# Spectroscopy of bright Algol-type semi-detached close binary system HU Tauri (HR 1471)

## M. Parthasarathy

### Indian Institute of Astrophysics, Bangalore – 560034, India

### Abstract

Spectroscopic results of HU Tauri that I presented in my PhD thesis (Parthasarathy 1979) are published in this paper. The H$\alpha$ line of the secondary is detected. The detection of H$\alpha$ line of the secondary is confirmed based on the high resolution Coude Reticon spectra obtained with the 2.1-m telescope of the McDonald observatory.

## 1. Introduction

The light variability of HU Tauri (HR 1471 = HD 29365, V = 5.92, Sp: B8V) was discovered by Strohmeier (1960). Strohmeier & Knigge (1960) found it to be an eclipsing binary with an orbital period of 2.056 days. Mammano & Margoni (1967) found the system to be a single lined spectroscopic binary. I made photometric and spectroscopic observations of this system and derived the photometric and spectroscopic elements and absolute dimensions of the components. The observational data and the results of the analysis were included in my Ph. D thesis (Parthasarathy 1979).

I found that the primary minimum to be an occultation eclipse wherein the B8V primary is eclipsed by the larger cool secondary component which has filled its Roche lobe. I have detected the H$\alpha$ line of the secondary component and from the radial velocities of the primary and secondary components the mass ratio is found to be 0.2564 (Parthasarathy 1979). Parthasarathy & Sarma (1980) published the B and V light curves of the system. Parthasarathy et al. (1993, 1995) derived the photometric elements using the Wilson & Devinney (1971) light curve synthesis method and confirmed the results obtained by Parthasarathy (1979). Tumer & Kurutac (1979), Dumitrescu & Dinescu (1980) and Dumitrescu & Suran (1993) also obtained the light curves of HU Tauri. Giuricin & Mardirossian (1981) analyzed the B and V light Curves of HU Tauri published by Parthasarathy and Sarma (1980). However their results were wrong because they assumed the primary minimum to be a transit. Ito (1988) has obtained complete B and V light curves; a solution to these light curves was presented by Nakamura et al. (1994). Maxted et al. (1995) obtained spectroscopic orbit and absolute parameters of HU Tauri which are in agreement with those obtained by Parthasarathy et al. (1993, 1995) and Parthasarathy (1979). In this paper I present the radial velocities, spectroscopic orbital elements and H$\alpha$ profiles of HU Tauri.

## 2.  Observations

Spectroscopic observations of HU Tauri in the blue and in the Hα region were made using the 102-cm telescope and Cassegrain spectrograph of the Kavalur observatory during the period January 1974 to December 1974.

All the spectra were obtained on photographic plates and were widened to 400 microns with a projected slit width of 20μ. A few spectra in the Hα region were widened to 800μ. The blue spectra were obtained on Eastman Kodak 103a–O and IIa-O (baked and unbaked) photographic plates. The spectra in the Hα region were obtained on Eastman Kodak 098-02, 103a-E and 103a–F photographic plates. Typical exposure times were thirty to sixty minutes for spectra in the blue and 90 minutes for spectra in the Hα region.

Fifty spectrograms in the blue region (25 Å/mm at Hγ) and twenty spectrograms in the Hα region (17 Å/mm) of HU Tauri were obtained. All spectra were measured with Zeiss Abbe comparator. The spectra in the blue cover a wavelength range from 3700 Å to 4500 Å. The spectral lines used for radial velocity measurement were all the Balmer lines. The He I 4026.2 Å and Si II 4128 Å lines were found to be very weak and were not used. Several radial velocity standard stars were observed. Radial velocities given in Tables 1 and 2 are on the standard system.

## 3.  Analysis

The columns in the Tables  1 and 2 give the plate number, the emulsion, the Heliocentric Julian day of the observation at mid–exposure, the phase, the measured radial velocity reduced to the sun (ref. Parthasarathy 1979 tables 9 and 10)  the results of the analysis are given in Tables 1, 2, 3 and 4 in this paper.

### Table 1. Radial velocities  of HU Tauri

| Plate No | Emulsion | JD(Hel) | Phase | Radial Velocity km/sec |
|----------|----------|---------|-------|------------------------|
| 1. | 2. | 3. | 4. | 5. |
| | | 2442000+ | | |
| 3142 | IIa-0 | 404.238 | 0.0042 | -17 |
| 3026 | " | 363.309 | 0.1054 | -41 |
| 3027 | " | 363.359 | 0.1295 | -62 |
| 3006 | 103a-0 | 361.312 | 0.1341 | -78 |
| 3111 | IIa-0 | 384.131 | 0.2313 | -67 |
| | | | | |
| 3112 | IIa-0 | 348.157 | 0.2439 | -59 |

| 2953 | " | 353.327 | 0.2512 | -54 |
|------|------|---------|--------|-----|
| 2520 | 103a-0 | 088.097 | 0.2668 | -63 |
| 3092 | IIa-0 | 382.233 | 0.3083 | -63 |
| 3093 | " | 382.268 | 0.3252 | -58 |
| | | | | |
| 3053 | 103a-0 | 378.206 | 0.3502 | -73 |
| 3164 | IIa-0 | 411.258 | 0.4186 | -24 |
| 3034 | " | 364.243 | 0.5598 | -04 |
| 3016 | " | 362.228 | 0.5795 | +00 |
| 2991 | 103a-0 | 360.242 | 0.6141 | +06 |
| | | | | |
| 2992 | " | 360.275 | 0.6298 | +16 |
| 3019 | IIa-0 | 362.441 | 0.6831 | +30 |
| 3137 | " | 389.298 | 0.7441 | +54 |
| 3100 | " | 383.143 | 0.7512 | +66 |
| 3138 | " | 389.321 | 0.7552 | +51 |
| | | | | |
| 3101 | II-a-O | 383.173 | 0.7656 | +62 |
| 3126 | " | 387.323 | 0.7838 | +62 |
| 3062 | " | 379.202 | 0.8342 | +40 |
| 3063 | " | 379.241 | 0.8528 | +42 |
| 3143 | " | 408.086 | 0.8759 | +43 |
| 3153 | " | 410.413 | 0.8762 | +21 |
| | | | | |
| **Radial velocities of HU Tauri derived from the Hα line of the primary** | | | | |
| | | | | |
| 3005 | 103a-E | 361.251 | 0.1044 | -30 |
| 3113 | " | 384.199 | 0.2642 | -67 |
| 2971 | 098.02 | 355.490 | 0.3030 | -63 |
| 2382 | " | 051.164 | 0.3056 | -60 |
| | | | | |
| 2396 | " | 053.225 | 0.3081 | -64 |
| 2494 | " | 086.413 | 0.3165 | -68 |
| 3122 | 103a-E | 387.132 | 0.6909 | +45 |
| 2995 | " | 360.426 | 0.7035 | +70 |
| | | | | |
| 2431 | 098.02 | 060.272 | 0.7349 | +52 |
| 2926 | " | 350.319 | 0.7884 | +68 |
| 2403 | " | 054.268 | 0.8153 | +46 |
| 3105 | 103a-E | 383.315 | 0.8346 | +56 |

### 3.1.1 The Hα line

The radial velocities of the primary component derived from the Hα absorption line are also given in Table 1 and they were also used in the orbit computation. A spectrogram (No.2382) obtained on 3[rd]

January 1974 shows a violet shifted broad emission feature (Figures 1 and 2). The peak velocity of the emission feature is found to be −600 km/sec (Figures 1 and 2). This spectrogram was obtained on Eastman Kodak 098-02 emulsion like rest of the H$\alpha$ plates. A few spectra in the H$\alpha$ region were obtained on 103a – E and 103a-F plates. The spectrogram of 3$^{rd}$ January is well exposed and it is widened to 800 microns and the exposure time was 89 minutes. The violet shifted emission feature extends very much in to the violet wing of the H$\alpha$ line. This emission feature is absent on a plate taken immediately after one orbital period. This indicates that this emission is a transient event. The same spectrogram shows absorption feature of the secondary towards the red side of the H$\alpha$ absorption core of the primary (Figure 2). The spectrum obtained on 6$^{th}$ January 1974 (plate No.2403, phase: 0.8153) shows clearly that this absorption feature is violet shifted with respect to the H$\alpha$ absorption core of the primary. This indicates that we are seeing the H$\alpha$ absorption line of the secondary.

### 3.1.2  The H$\alpha$ line of the secondary

The radial velocities of the secondary component derived from its H$\alpha$ line are given Table 2 (ref. Parthasarathy 1979, table 10**).**  The H$\alpha$ line of the secondary of HU Tauri is clearly seen in the high resolution Coude Reticon spectra of HU Tauri obtained with the 2.1-m Otto Struve telescope of the McDonald observatory (Figure 3).

From the radial velocities of the H$\alpha$ line of the secondary (Table 2) K2 is found to be +234 km/sec. The mass ratio m2/m1 = K1/K2 is found to be 60/234 = 0.2564. Figure 3 shows the high resolution H$\alpha$ region spectra obtained on 1981 December 18$^{th}$ (phase = 0.2402), on February 28$^{th}$ (phase = 0.7579) and at phase 0.9833 on 1982 February 17$^{th}$. The H$\alpha$ lines of the primary and secondary are relatively broad indicating that they are rotating rapidly.

**Table 2.  Radial Velocities derived from the H$\alpha$ line of the secondary**

| Plate No. | Emulsion | JD (hel) | Phase | Velocity km/sec |
|-----------|----------|----------|-------|-----------------|
|           |          | 2442000+ |       |                 |
| 3008      | 098-02   | 361.437  | 0.1949 | -              |
| 3113      | 103-aE   | 384.199  | 0.2642 | +273           |
| 2935      | 098-02   | 351.324  | 0.2769 | +243           |
| 2382      | "        | 051.164  | 0.3056 | +219           |
| 2396      | "        | 053.225  | 0.3081 | +240           |
|           |          |          |        |                |
| 2494      | "        | 086.143  | 0.3165 | +223           |
| 3017      | 103-aF   | 362.306  | 0.6177 | -              |
| 2431      | 098-02   | 060.272  | 0.7349 | -              |
| 2926      | "        | 350.319  | 0.7884 | -              |
| 2403      | "        | 054.268  | 0.8153 | -208           |

**Table 3.  Spectroscopic orbital elements of HU Tauri**

| Vo | -6.5 km/sec |
|----|-------------|
| K1 | 60.0 km/sec |
| K2 | 234.0 km/sec |
| K1/K2 | 0.2564 |
| e | 0.0 |
| a1sini | $1.781 \times 10^6$ km |
| a2sini | $6.622 \times 10^6$ km |
| m1sin$^3$i | 4.42 M$_\odot$ |
| m2sin$^3$i | 1.19 M$_\odot$ |

   The probable errors in Vo, K1, K2 are found to be  2 km/sec, 2.5 km/sec and 3.5 km/sec respectively.

## 4.  Conclusions

The photometric, spectroscopic elements and absolute dimensions derived by Parthasarathy (1979) are in good agreement with those derived by Parthasarathy et al. (1993, 1995), Ito (1988), Nakamura et al. (1994) and Maxted et al. (1995). The masses and radii derived by me and Maxted et al. (1995) are given in Table 4.

Table 4.  Masses and radii of the components of HU Tauri

|  | Parthsarathy et al. (1979) | Parthsarathy et al. (1995) | Maxted et al. (1995) |
|---|---|---|---|
| M1 | 4.8 M$_\odot$ | 4.68 M$_\odot$ | 4.43 M$_\odot$ |
| R1 | 2.7 R$_\odot$ | 2.90 R$_\odot$ | 2.57 R$_\odot$ |
| M2 | 1.3 M$_\odot$ | 1.26 M$_\odot$ | 1.14 M$_\odot$ |
| R2 | 3.5 R$_\odot$ | 3.34 R$_\odot$ | 3.21 R$_\odot$ |

The H$\alpha$ line of the secondary detected on photographic plates is confirmed with the high resolution Coude Reticon spectra of HU Tauri obtained with the 2.1-meter Otto Struve telescope of the McDonald Observatory (Figure 3). The strength of the H$\alpha$ line of the secondary (Figures 1 and 3) indicates that it may be a late F – early G III-IV type star.

HU Tauri is a semi-detached Algol type close binary system. The primary minimum in the light curve is due to an occultation eclipse. The secondary has filled its Roche lobe and mass-transfer and gaseous streams seem to be present in the system, the phase interval 0.56 to 0.68 seems to be affected. Maxted et al. (1995) also mention that around phase 0.15 there is some scatter. In the IUE UV high resolution spectrum of HU Tauri outside the eclipse Si IV (1393.755 Å, 1402.770 Å) absorption feature is found, which indicates the presence of high temperature plasma between the components or close to the B8V primary.

Further study of the system based on high resolution and high signal to noise ratio spectra is needed.

### Acknowledgements

I am very much thankful to late Prof. M. K. V Bappu for generously allotting observing time on the 1-meter telescope of the Kavalur observatory. I am also very much thankful to late Prof. Harlan J Smith for generously allotting observing time on the 2.1 m Otto Struve telescope of the McDonald observatory.

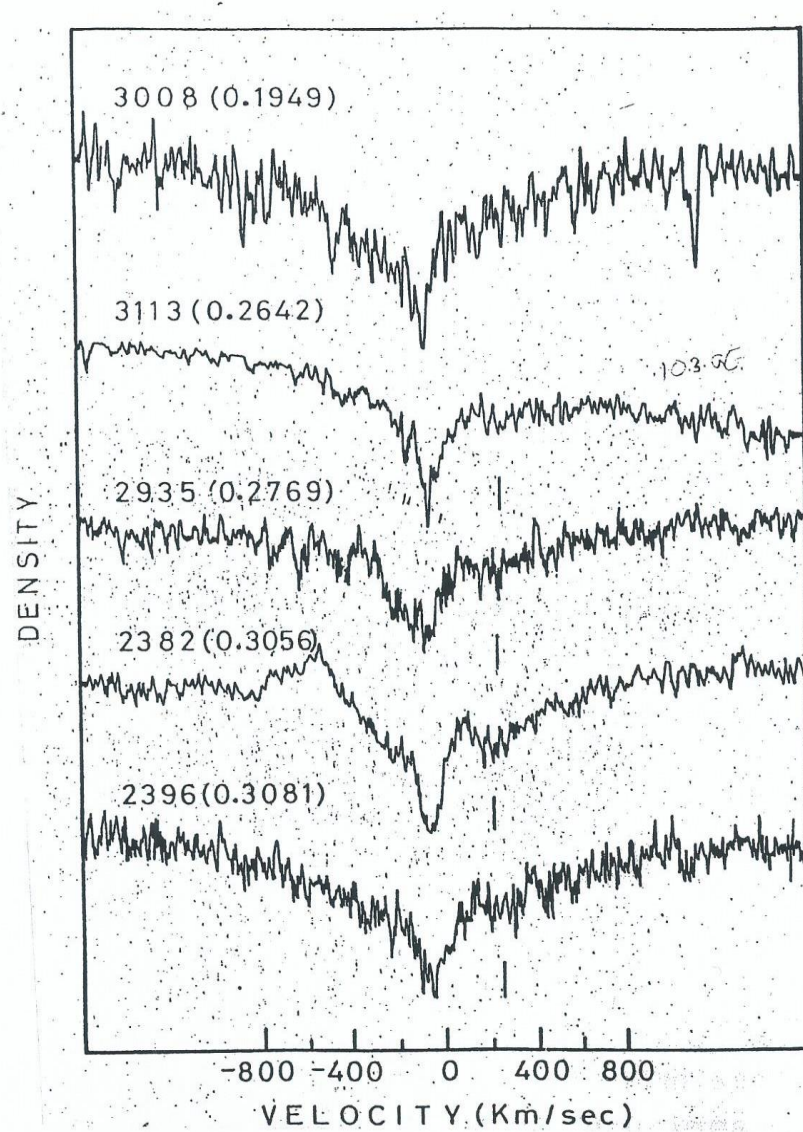

Figure 1 . The Hα profiles of HU Tauri at different phases. The Hα absorption line of the secondary is marked in the figure. Plate numbers and phases are given in the figure.

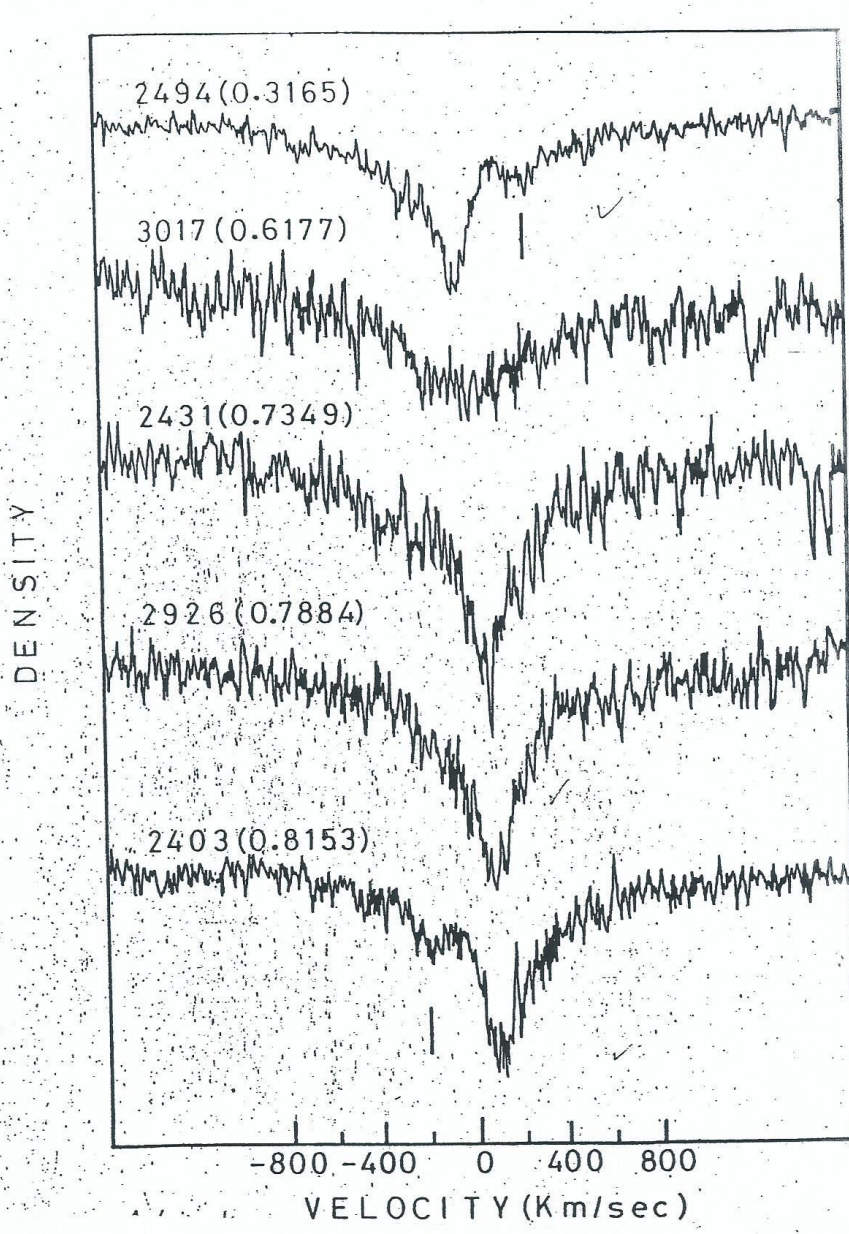

Figure 1. (continued)

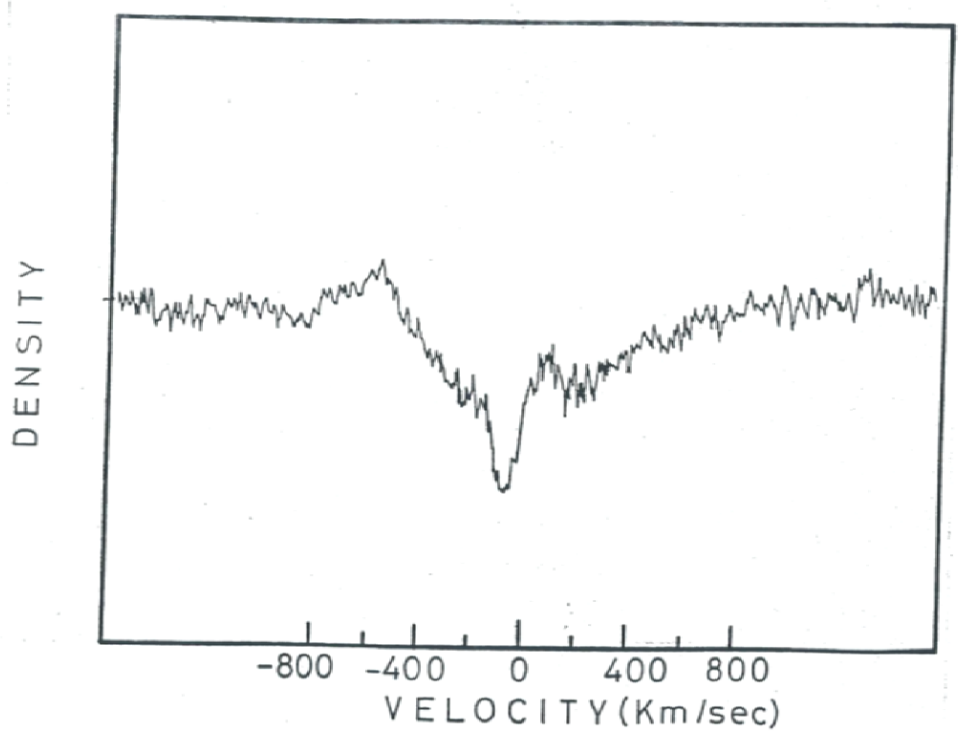

Figure 2 . A spectrogram (No .2382) obtained on 3[rd] January 1974 (phase : 0.3056) shows a violet shifted broad emission feature. The peak velocity of the emission feature is found to be -600 km/sec. The Hα absorption feature of the secondary is marked in the figure.

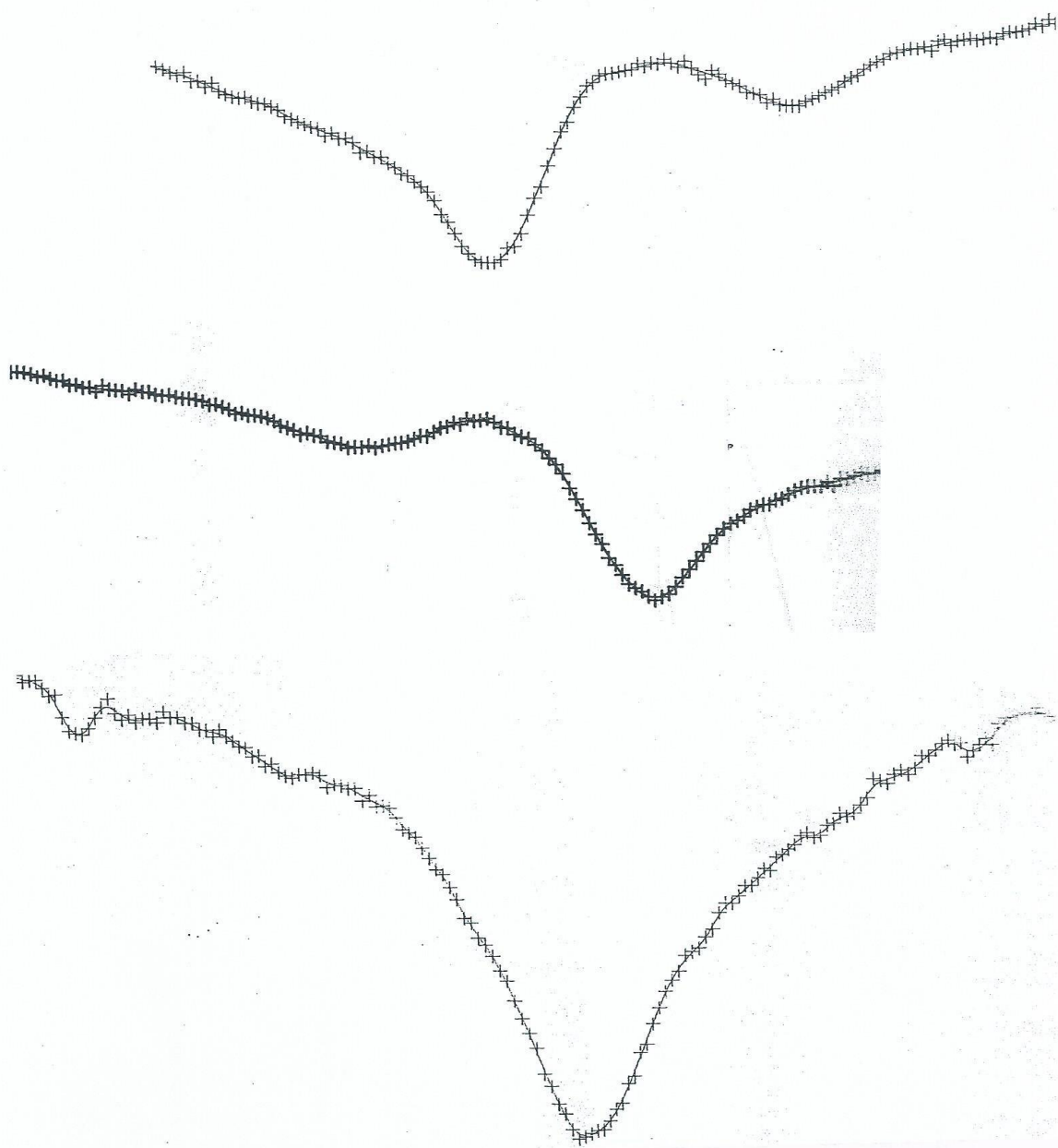

Figure 3. Coude Reticon high resolution spectra of HU Tauri in the Hα region obtained with the 2.1-m Otto Struve telescope of the McDonald observatory. The Hα line of the secondary is marked. The Hα absorption lines of the primary and secondary at phase 0.2402 are clearly seen. Top : phase 0.2402 (1981 December 18[th]), middle : phase 0.7579 (1983 February 28[th]), bottom : phase 0.9833 (1982 February 17[th]).